\documentclass[12pt]{iopart}
\usepackage{iopams,epsf}

\usepackage{color}
\usepackage{epsfig}
\usepackage{graphics}
\usepackage{psfrag}
\usepackage{lscape}

\begin{document}

\title[Microcanonical entropy for small magnetisations]{Microcanonical
entropy for small magnetisations}

\author{Hans Behringer}

\address{Institut f\"ur Theoretische Physik I, Universit\"at Erlangen-N\"urnberg, D -- 91058 Erlangen, Germany}

\ead{Hans.Behringer@physik.uni-erlangen.de }

\begin{abstract}
Physical quantities obtained from the microcanonical entropy
surfaces of classical spin systems show typical features of phase
transitions already in finite systems. It is demonstrated that the
singular behaviour of the microcanonically defined order parameter
and susceptibility can be understood from a Taylor expansion of
the entropy surface. The general form of the expansion is
determined from the symmetry properties of the microcanonical
entropy function with respect to the order parameter. The general
findings are investigated for the four-state vector Potts model as
an example of a classical spin system.
\end{abstract}




\section{Introduction}

The macroscopic behaviour of a physical system in thermodynamic
equilibrium is related to the microscopic properties by
statistical mechanics. The basic quantity in this connection is
the density of states depending on the macroscopic quantities of
interest. For a classical spin system as a model system to
describe magnetic properties these macroscopic variables are the
energy and the magnetisation. In the traditional approach to the
statistical description of phase transitions the density of states
is Laplace-transformed to the partition function which is a
concept of the canonical ensemble and the physics is deduced from
the corresponding potential, the Gibbs free energy. In recent
years an alternative approach to analyse phase transitions within
a statistical framework has been developed in the microcanonical
ensemble (Gross, 1986a; 1986b; 2001; H\"uller, 1994;
Promberger/H\"uller, 1995; Gross et al., 1996). Apart from works
about general questions of the equivalence and inequivalence of
the various ensembles (Lewis et al., 1994; Dauxois et al., 2000;
Barr\'{e} et al., 2001; Ispolatov/Cohen, 2001) and of the
thermodynamic limit in the microcanonical ensemble ((Kastner 2002)
to cite a recent work) also second order phase transitions have
been studied recently (Kastner et al., 2000; H\"uller/Pleimling,
2002). Ways to extract critical exponents from microcanonical
quantities have been suggested (Kastner et al., 2000;
H\"uller/Pleimling, 2002). In these works is was numerically
demonstrated that the microcanonically defined physical quantities
like the microcanonical order parameter or the susceptibility
exhibit singularities already in finite systems. However, these
numerical studies suggest that the microcanonical singularities
are characterised by classical critical exponents in contrast to
the non-trivial exponents showing up in the thermodynamic limit.

The aim of the present paper is to provide an analytic understanding of the
characterisation of the singularities in physical quantities of
finite microcanonical systems by classical exponents. To this end
the entropy surface is expanded into a Taylor series in terms of
its natural variables energy and magnetisation   and the
asymptotic behaviour of the macroscopic quantities is
investigated. This procedure is similar to the Landau expansion of
the free energy of the infinite system although a striking
difference has to be stressed. The Landau approach is an {\em
approximation} of the free energy of the {\em infinite} canonical
system whereas the treatment presented below is {\em exact} for
the {\em finite} microcanonical system.

The work is organised in the following way. The second section
contains a short survey of the microcanonical approach to
statistical mechanics. In particular the basic concepts and
findings of the investigation of second order phase transitions
are summarised. In the third section the appearance of the
classical critical exponents is related to the analyticity of
the thermodynamic potential of finite systems. The fourth section
contains general considerations about the Taylor expansion of
entropy surfaces with a $C_{nv}$ order parameter symmetry. In
addition an investigation of the entropy surface of
finite vector Potts models with four states is presented. The
analysed data are obtained from Monte Carlo simulations.

\section{Thermostatics in microcanonical systems}

The analysis of statistical properties of classical spin systems within the microcanonical
ensemble starts from the density of states
\begin{equation}
    \label{eq:def_dos}
    \Omega(E, M, L^{-1}) = \sum_{\sigma \in \Gamma_N} \delta_{E, \mathcal{H}(\sigma)}
    \delta_{M, \mathcal{M}(\sigma)} \;.
\end{equation}
Here it is assumed that the system is defined on a $d$-dimensional
hypercube of extension $L$. The Hamiltonian $\mathcal{H}$ gives
the total energy of the microstate $\sigma$ and the magnetisation
operator $\mathcal{M}$ its total magnetisation. In general the
magnetisation is a multicomponent object $M = (M_1, \ldots, M_n)$.
The microcanonical entropy is obtained by taking the logarithm of
the density of states:
\begin{equation}
    s(e,m, L^{-1}) = \frac{1}{L^d} \ln \Omega(E,M, L^{-1})\;.
\end{equation}
To compare different system sizes the extensive factor $L^d$ is
divided out to give the specific energy $e=E/L^d$ and the specific
magnetisation $m=M/L^d$. The dependence of the physical quantities
on the system size $L$ is suppressed in the following. The
statistical properties of an isolated physical system are deduced
from the entropy surface. The microcanonical spontaneous
magnetisation of a finite system for a fixed energy is defined by
\begin{equation}
    \label{eq:def_spontane_mag}
    m_{\scriptsize \mbox{sp}}(e) : \iff s(e, m_{\scriptsize \mbox{sp}}(e)) = \max_{m} s(e, m)  \;,
\end{equation}
hence $m_{\scriptsize \mbox{sp}}(e)$ corresponds to the magnetisation with the
maximum entropy for a given energy.

The non-vanishing multicomponent spontaneous magnetisation vector
defines a direction in the order parameter space:
\begin{equation}
    \label{eq:zerlegung_ord}
    m_{\scriptsize \mbox{sp}}(e) = (m_{\scriptsize \mbox{sp}}^{(1)}(e), \ldots, m_{\scriptsize \mbox{sp}}^{(n)}(e)) =
    |m_{\scriptsize \mbox{sp}}(e)| m_0
\end{equation}
where $m_0$ is a fixed unit vector with several possible
orientations below $e_{\scriptsize \mbox{c}}$. Note however that it might
be possible that $m_0$ and consequently the associated
orientations are themselves functions of the energy below $e_{\scriptsize
\mbox{c}}$. Let $G_0$ be the symmetry group of the microcanonical
entropy with respect to the magnetisation components, i.e.
\begin{equation}
    \label{eq:allg_sym}
    s(e, g(m)) = s(e, m)
\end{equation}
for all transformations $g$ in $G_0$. If the spontaneous
magnetisation $m_{\scriptsize \mbox{sp}}$ is finite the symmetry group $G_0$ of the
entropy is broken down to the subgroup $G < G_0$ that leaves the
direction $m_0$ invariant (Behringer, 2003).

In the microcanonical ensemble of isolated systems no external
magnetic field appears. However, magnetic fields can be defined as
the conjugated variable of the magnetisation components giving
rise to the relation\footnote{The sloppy but
more convenient notation $\frac{\partial}{\partial x} f(x_0)$ for
$\frac{\partial}{\partial x} f(x)|_{x=x_0}$ is used in the
following, upper indices are not discriminated from lower ones.}
\begin{equation}
\label{eq:def_magnetfeld}
    h_i(e,m_{\scriptsize \mbox{sp}}(e))  \frac{\partial}{\partial e}
    s(e,m_{\scriptsize \mbox{sp}}(e))=
-\frac{\partial}{\partial
    m_i} s(e,m_{\scriptsize \mbox{sp}}(e))
\end{equation}
for the $i$th component in equilibrium.

The susceptibility of the system is related to the curvature of
the entropy along the magnetisation direction. For a
multicomponent magnetisation the curvature at the spontaneous
magnetisation is related to the Hessian
\begin{equation}
H_{kl}(e,m_{\scriptsize \mbox{sp}}(e)) = \frac{\partial^2}{\partial m_k \partial
m_l}s(e, m_{\scriptsize \mbox{sp}}(e))
\end{equation}
of the entropy and the susceptibility is consequently a tensor
which is given by
\begin{equation}
    \label{eq:sus_gleichgewicht}
    \chi^{(0)}_{ij}(e) = -\frac{\partial}{\partial e} s(e, m_{\scriptsize \mbox{sp}}(e))
    \left(H^{-1}(e,m_{\scriptsize \mbox{sp}}(e))\right)_{ij} \;.
\end{equation}

From a geometrical point of view the specific heat of the system is connected to the curvature along the
energy direction. At the spontaneous magnetisation the specific heat is
given by
\begin{equation}
\label{eq:definition_spezifische}
c^{(0)}(e) = -\frac{\left(\frac{\partial}{\partial e}s(e,m_{\scriptsize
\mbox{sp}}(e))\right)^2}{\frac{\partial^2}{\partial e^2} s(e,m_{\scriptsize \mbox{sp}}(e))}\;.
\end{equation}
Alternatively one can consider a specific heat that is more directly related
to the canonical viewpoint. There the specific heat is calculated from the
canonical entropy 
\begin{equation}
s_{\scriptsize \mbox{can}} = -\frac{\partial}{\partial T} g(T,h)
\end{equation}
obtained from the Gibbs free energy $g(T,h)$ of the system in equilibrium as
\begin{equation}
 c_{\scriptsize \mbox{can}} = T\frac{\partial}{\partial T} s_{\scriptsize \mbox{can}}\;.
\end{equation}
Defining the equilibrium entropy by
\begin{equation}
\tilde{s}(e) = s(e,m_{\scriptsize \mbox{sp}}(e))
\end{equation}
one obtains the alternative specific heat
\begin{equation}
\label{eq:alternative_spez} \tilde{c}(e) =
-\frac{\left(\frac{\mbox{d}}{\mbox{d}
e}\tilde{s}(e)\right)^2}{\frac{\mbox{d}^2}{\mbox{d} e^2} \tilde{s}(e)}\;.
\end{equation}
The specific heat $\tilde{c}$ is different from $c^{(0)}$ as
$m_{\scriptsize \mbox{sp}}$ is first plugged into $s$  and the derivative
with respect to the energy is worked out afterwards. In contrast,
$c^{(0)}$ is evaluated according to footnote $\ddagger$.


From a statistical point of view a phase transition from a
disordered high-symmetric macrostate to an ordered low-symmetric
macrostate is defined to take place at a non-analytic point of the
corresponding thermostatic potential describing the properties of
the physical system. Yang and Lee showed that the grand-canonical
thermostatic potential, i.e. the Gibbs free energy, is an analytic
function for all finite system sizes (Yang/Lee, 1952; Lee/Yang,
1952). This means that a phase transition can only occur in the
thermodynamic limit of an infinite system, according to the
definition above phase transitions are not possible in finite
systems.

However the microcanonical equilibrium quantities exhibit typical
features of phase transitions (Kastner et al., 2000). For instance
the microcanonical spontaneous magnetisation
(\ref{eq:def_spontane_mag}) displays the characteristics of
spontaneous symmetry breaking. This spontaneous magnetisation
defines the order parameter. For the high energy phase the
microcanonical order parameter is zero reflecting a phase with
high symmetry. The abrupt emerging of a finite order parameter for
low energies indicates the transition to an ordered phase with
lower symmetry. The energy at which this onset occurs defines
unambiguously the critical energy $e_{\scriptsize \mbox{c}}$ of the finite
microcanonical system. The susceptibility
(\ref{eq:sus_gleichgewicht}) of the system diverges at this
energy.

The use of the expressions {\em critical} and {\em phase
transition} to describe the behaviour of finite microcanonical
systems is somewhat problematic as they are commonly used for
certain properties of the infinite system. Nevertheless the
appearance of features of the microcanonical quantities which are
also found in infinite systems at the transition point suggests
its usage. Another delicate point is the discreteness of the
physical quantities of discrete spin systems. There the used
language refers to continuous functions that describe the discrete
data most suitably. In the microcanonical analysis of continuous
spin systems as e.g. the $xy$ (Richter et al., 2004) or the
Heisenberg model this concern does not exist.

The existence of a nontrivial spontaneous magnetisation is in
general related to the appearance of a convex dip in the
microcanonical entropy. This is different in the canonical
ensemble where the curvature along the natural variables is
connected to the mean square deviation of a physical quantity. The
curvature has therefore a well-defined sign. Although the physical
quantities show singularities at the transition point that are
typical of phase transitions there is no phase transition as long
as the microcanonical system is finite. The microcanonical entropy
of finite systems is an analytic function of its natural variables
and hence a phase transition in the above defined mathematical
sense does not take place.

The singular behaviour of the order parameter with respect to the
energy can, however, be described by means of a critical exponent
(Kastner et al., 2000). For a physical quantity $Q(e)$ of the finite system this critical
exponent, say $\tilde{\kappa}$, is defined by
\begin{equation}
\tilde{\kappa} = \lim_{e \to e_{\scriptsize \mbox{c}}} \frac{\mbox{d}
\ln|Q(e)|}{\mbox{d} \ln|e-e_{\scriptsize
\mbox{c}}|}\;.
\end{equation}
The microcanonical critical exponent $\tilde{\beta}$ of the
spontaneous magnetisation in finite systems turns out to have the
classical value 1/2, the corresponding exponent $\tilde{\gamma}$
of the susceptibility is 1 for all system sizes. Both definitions
(\ref{eq:definition_spezifische})  and (\ref{eq:alternative_spez})
of the specific heat are characterised by the exponent
$\tilde{\alpha} = 0$. Whereas the specific heat $c^{(0)}$ has a kink at
the critical energy, $\tilde{c}$ exhibits a jump at $e_{\scriptsize
\mbox{c}}$ with different values $\tilde{c}(e_c^+) \neq
\tilde{c}(e_c^-)$. The kink of $c^{(0)}$ at the critical energy with
$c^{(0)}(e_c^+)= c^{(0)}(e_c^-)$, but different
left-handed and right-handed derivatives at $e_c$, however, is not
a cusp-singularity according to the definition of (Stanley, 1972)
as the higher order derivatives of $c^{(0)}$ do not diverge at $e_{\scriptsize
\mbox{c}}$. Note also that both quantities tend to the same function
in the thermodynamic limit.

\section{Taylor expansion of the microcanonical entropy}
\label{kap:taylor}

For a finite system with the typical characteristics of a
continuous phase transition the microcanonically defined order
parameter near the critical point vanishes like a square root.
This behaviour can be understood in a Landau type of approach to
the description of phase transitions in the microcanonical
ensemble. To this end the entropy is expanded into a Taylor series
as a function of the order parameter components. A proof of the
analyticity of the microcanonical entropy is still lacking.
Nevertheless in view of the analyticity of the canonical
thermostatic potential it seems obvious to use this assumption.
Note that this approach is exact provided the entropy of a finite
system is an analytic function and hence can be expanded about any
point $(e,m)$ in the parameter space. This is different in the
Landau theory of the free energy of the infinite system. In this
approximation the Helmholtz free energy is assumed to be analytic
although a phase transition in the thermodynamic limit is related
to a non-analytic point in the thermostatic potential.
Consequently the critical behaviour obtained from the Landau
approximation may differ from the true non-trivial behaviour of
the physical systems. This is indeed the case below the upper
critical dimension.

The Taylor expansion of the microcanonical entropy of the finite
system about the magnetisation $m = 0$ (the equilibrium
magnetisation of the high energy phase) yields the general series
\begin{equation}
    \label{eq:allg_entwicklung}
    s(e, m) = s(e, m=0) + \sum_{(p_1, \ldots, p_n)} a_{p_1, \ldots, p_n}
    (e) \prod_{i}m_i^{p_i} \;,
\end{equation}
where the expansion coefficients are related to the derivatives of
the entropy with respect to the magnetisations
\begin{equation}
    a_{p_1, \ldots, p_n}(e) = \frac{1}{p_1! \ldots p_n!}\partial^{p_1}_{m_1} \ldots \partial^{p_n}_{m_n}
    s(e, m=0) \;.
\end{equation}
The symmetry group $G_0$ of the entropy with respect to the
magnetisation determines which coefficients $a_{p_1, \ldots, p_n}$
appear in the expansion (\ref{eq:allg_entwicklung}). Only the
coefficients which correspond to monomials $m_1^{p_1} \ldots
m_n^{p_n}$ that are invariant under the transformations $g \in
G_0$ are in general non-zero, all other coefficients necessarily
vanish. In view of the symmetry $G_0$ the expansion
(\ref{eq:allg_entwicklung}) is in fact an expansion in terms of
the symmetry-adapted harmonics of the group $G_0$.

\begin{figure}
\begin{center}
\epsfig{file=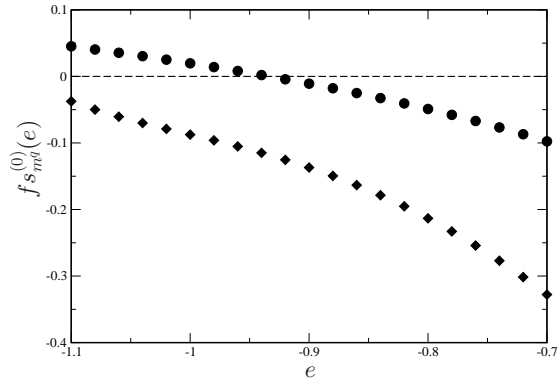,width=13em, angle=270}
\end{center}
\caption{\label{schieftaylorkoef} \small The derivatives
$\partial_{m^2} s(e,0)$ and $\partial_{m^4} s(e,0)$ of the entropy
surface of a two-dimensional Ising model with $200$ spins in a
square lattice geometry. The fourth order derivative is multiplied
by a factor $f=0.1$. The second order derivative changes its sign
at the transition energy $e_{\scriptsize \mbox{c}}=-0.934$ whereas the
fourth order derivative stays negative in the vicinity of $e_{\scriptsize
\mbox{c}}$.}
\end{figure}

The physical behaviour for small magnetisations is already
described by the low order terms in (\ref{eq:allg_entwicklung}).
Suppose the magnetisation is only one-dimensional and the
microcanonical entropy is left invariant by the transformation $m
\to -m$. Note that these are the symmetry properties of the
entropy of the Ising model. Then only even powers contribute to
the Taylor expansion and up to the fourth order term  one gets
\begin{equation}
    \label{eq:landau_ord4}
    s(e,m) = s(e, 0) + \frac{1}{2} a(e)m^2 + \frac{1}{4} b(e) m^4.
\end{equation}
As the maximum of the entropy must not be at an infinite
magnetisation the coefficient of the highest power of $m$ has to
be negative. This requirement is indeed fulfilled for an Ising
model in the energy interval about $e_{\scriptsize \mbox{c}}$ where a
finite order parameter emerges. The Ising model is defined by the
Hamiltonian
\begin{equation}
\mathcal{H} = -\sum_{\left<ij\right>} \sigma_i\sigma_j
\end{equation}
where the spins $\sigma_i$ can be in the state $+1$ or $-1$ and the summation
runs over neighbour pairs of spins.
Figure \ref{schieftaylorkoef}
displays the derivatives $\partial_{m^q} s(e,0)$ of a
two-dimensional Ising model with 200 spins near the energy
$e_{\scriptsize \mbox{c}}$. The underlying two-dimensional lattice has a
square-lattice topology. The density of states of the system is
evaluated numerically exact by a microcanonical transfer matrix
method (Binder, 1972; Creswick, 1995) so that it is possible to
work out higher order derivatives reliably.
The condition (\ref{eq:def_spontane_mag}) for
thermostatic equilibrium
leads to the equation
\begin{equation}
    \frac{\partial}{\partial m} s(e,m) = m \left(a(e) + b(e) m^2 \right) \stackrel{}{=} 0 \;.
\end{equation}
for $m_{\scriptsize \mbox{sp}}(e)$. This relation has  always the trivial solution
$m_{\scriptsize \mbox{sp}}(e) = 0$ and in addition the solutions
\begin{equation}
    \label{eq:mag_nichttrivial}
    m_{\scriptsize \mbox{sp}} (e) = \pm \sqrt{- \frac{a(e)}{b(e)}}
\end{equation}
with a non-trivial, finite order  parameter if the coefficient
$a(e)$ is positive. The stability condition
\begin{equation}
    \frac{\partial^2}{\partial m^2} s(e,m_{\scriptsize \mbox{sp}}(e)) \stackrel{}{<} 0
\end{equation}
is always satisfied for the non-trivial  solution
(\ref{eq:mag_nichttrivial}) provided it exists. The solution
$m_{\scriptsize \mbox{sp}}(e) = 0$ is only stable, i.e. corresponds to a
maximum of the entropy, if the coefficient $a(e)$ is negative. The
coefficient $a(e)$ changes its sign at the energy $e_{\scriptsize
\mbox{c}}$. Hence the equilibrium magnetisation of the entropy
(\ref{eq:landau_ord4}) is given by
\begin{equation}
    \label{eq:mag_landau}
    m_{\scriptsize \mbox{sp}}(e) = \pm \Theta(a(e)) \sqrt{\left|
\frac{a(e)}{b(e)}\right|} \;,
\end{equation}
where the Heaviside function is denoted by $\Theta$. As the
coefficient $b(e)$ is negative in the vicinity of $e_{\scriptsize
\mbox{c}}$ and the two functions $a$ and $b$ are smooth the order
parameter varies continuously as a function of the energy $e$.
Below the energy $e_{\scriptsize \mbox{c}}$ it attains a finite value. Note
that at the energy $e_{\scriptsize \mbox{c}}$ the solution $m = 0$ becomes
instable and the curvature of the entropy along the magnetisation
changes its sign. The existence of two stable solutions below
$e_{\scriptsize \mbox{c}}$ signals the spontaneous breakdown of the global
$m \to \pm m$ symmetry of the physical system. The depth
\begin{equation}
    \Delta(e) = s(e, m_{\scriptsize \mbox{sp}}(e)) - s(e, 0)
\end{equation}
of the convex dip along the magnetisation for fixed energy can
also be expressed in terms of the expansion coefficients  of the
entropy surface. Near the critical point of the finite system one
obtains the approximation
\begin{equation}
    \label{eq:durch_landau}
    \Delta(e) \approx - \Theta(a(e)) \frac{a^2(e)}{4b(e)} \;.
\end{equation}

The Landau expansion of the entropy surface around the equilibrium
magnetisation of the high energy phase provides an explanation of
the classical exponents describing the singular behaviour of the
microcanonical physical quantities of the finite system.
Introducing the reduced energy
\begin{equation}
    \varepsilon := e - e_{\scriptsize \mbox{c}}
\end{equation}
one can expand the functions  $a(e)$ and $b(e)$ about the critical energy $e_{\scriptsize \mbox{c}}$:
\begin{equation}
    a(e) = -A\varepsilon + a_2 \varepsilon^2 + \ldots
\end{equation}
and
\begin{equation}
    b(e) = -B + b_1 \varepsilon + b_2 \varepsilon^2 + \ldots
\end{equation}
where the constants $A$ and $B$ are positive. Plugging these
expansions into relations (\ref{eq:mag_landau}) and
(\ref{eq:durch_landau}) one gets the leading behaviour
\begin{equation}
    m_{\scriptsize \mbox{sp}}(\varepsilon) = \pm \Theta (-\varepsilon) \sqrt{\frac{A}{B}} \sqrt{|\varepsilon|}
\end{equation}
for the order parameter and
\begin{equation}
    \Delta (\varepsilon) = \Theta (-\varepsilon) \frac{A^2}{4B} \varepsilon^2
\end{equation}
for the depth of the convex dip in the limit of small reduced
energies $\varepsilon$. Thus the variation of the spontaneous
magnetisation in finite systems is described by the classical
critical exponent $\tilde{\beta} = 1/2$. In addition one can
deduce the leading behaviour of the curvature of the entropy along
the magnetisation. For energies above the critical point, i.e.
$\varepsilon > 0$ one has
\begin{equation}
    \partial_{mm} s(\varepsilon, m_{\scriptsize \mbox{sp}}(\varepsilon)) = -A \varepsilon
\end{equation}
and for negative reduced energies one has
\begin{equation}
    \partial_{mm} s(\varepsilon, m_{\scriptsize \mbox{sp}}(\varepsilon)) = 2A \varepsilon \;.
\end{equation}
This leads to the critical exponent $\tilde{\gamma}=1$
characterising the divergent susceptibility in finite
microcanonical systems. Note that the slope of the curvature
$\partial_{mm} s$ at the spontaneous magnetisation above and below
the critical point differ by a factor two. This is also observed
in the curves obtained from the numerically calculated entropy
surface (see discussion below and figure \ref{bildsmm_z4}). From
the definition (\ref{eq:def_magnetfeld}) of the magnetic field and
the expansion (\ref{eq:landau_ord4}) it is apparent that the
magnetic field at the critical energy $e_{\scriptsize \mbox{c}}$ vanishes
with an exponent $\tilde{\delta} = 3$ in the limit of small
magnetisations. Similarly the critical exponent $\tilde{\alpha}$ can be
shown to be $0$ for both definitions of the specific heat (compare (\ref{eq:definition_spezifische})  and
(\ref{eq:alternative_spez})).

The expansion of the entropy surface hence describes the
asymptotic behaviour of the physical quantities in the vicinity of
the critical point of the finite system. It has to be stressed
once again that the expansion of the entropy surface is not an
approximation like the Landau expansion of the Helmholtz free
energy. The Taylor expansion of the microcanonical entropy about
an arbitrary point in the $(e, m)$ space is always possible due to
the analyticity of the entropy of finite systems. For the infinite
system such an expansion about the critical point can not be
performed as the thermostatic potential is singular precisely at
this point. Therefore non-trivial critical exponents can only
emerge in the thermodynamic limit.

Although the physical quantities of finite microcanonical systems
are always characterised by classical exponents precursors of the
non-trivial exponent of the infinite system show up if the
evolution of the quantities is considered for a series of
different system sizes. This evolution of the physical quantities
with increasing size $L$ can be investigated with effective
critical exponents (H\"uller/Pleimling, 2002) or microcanonical
finite size scaling relations (Kastner et al., 2000). The
microcanonical finite size scaling theory is conceptionally
similar to the corresponding finite size scaling theory of the
canonical ensemble.

At this stage one delicate point has to be stressed again. Whereas
the Taylor expansion of the entropy can be carried out without any
ambiguity for systems with continuous energies and magnetisations
(e.g. $xy$ model) this seems to be doubtful for discrete models
such as the Ising model. A suitable chosen continuous function may
be fitted to the data. This leads, however, to a certain degree of
arbitrariness which is still related to the choice of this
function. Alternatively, differentials can be replaced by (centre)
differences leading to a discrete set of data points.
Nevertheless, any properly  chosen function will reproduce the
change of the curvature along the magnetisation at $m=0$ and
$e_{\scriptsize \mbox{c}}$, leading to maxima at non-zero
magnetisations. It is precisely this bifurcation at
$e_{\scriptsize \mbox{c}}$ that gives rise to classical exponents.
Different fit-functions will only lead to slightly different
coefficients, the overall dependence of the expansion an $e$ and
$m$  will be the same. The considerations in this section strictly
valid only for systems with continuous $e$ and $m$ give a
heuristic understanding of the classical behaviour of physical
quantities of discrete systems.

\section{Entropy surface with $C_{nv}$ symmetry}

\subsection{Landau expansion of entropies with $C_{nv}$ invariance}

A general method that allows the determination of the invariant
homogeneous polynomials which can appear in the Taylor expansion
of the entropy is based on the concept of the complete rational
basis of invariants of a group. The determination of the Landau
expansion of the free energy of an infinite system with the help
of the complete rational basis of invariants was first proposed by
Gufan in (Gufan, 1971) (see also Tol\'edano/Tol\'edano, 1987). In
this subsection this general method is applied to a microcanonical
entropy surface with a $C_{nv}$ symmetry. A prominent example of a
physical system with this symmetry property of the entropy surface
is the $n$-state vector Potts model which will be investigated in
section \ref{kap:vierpotts} for a two-dimensional lattice.

An invariant homogeneous monomial of degree $p$ with respect to a
finite group can be constructed as a linear combination of
products of a limited number of polynomials forming the complete
rational basis of invariants. In polar coordinates $m_1 =
r\cos\phi$ and $m_2 = r\sin \phi$ the complete rational basis of
invariants of the finite group $C_{nv}$ consists of the two
polynomials $r^2$ and $r^n\cos n \phi$. The polynomial
$r^2=m_1^2+m_2^2$ is always invariant under the modulus preserving
transformations of the
group $C_{nv}$.

The entropy of a finite microcanonical system with the symmetry
\begin{equation}
    \label{eq:entro_symmetrie_c4n}
    s(e, m_1, m_2) = s(e, C_{nv}(m_1, m_2))
\end{equation}
can be expanded into a Taylor series. In polar coordinates the
polynomials appearing in the expansion are sums of products of the
monomials of the rational basis of invariant of $C_{nv}$. The
basic monomials contain the angular dependence of the
magnetisation exclusively in the form $\cos n\phi$. Thus the
entropy in polar coordinates is of the form
\begin{equation}
    s(e, m_1, m_2) = \tilde{s}(e, r, \cos n\phi) \;.
\end{equation}
This form of the entropy already determines properties that are
independent of the details of the Taylor expansion, as e.g. the
highest degree in the truncated Taylor series. These properties
are solely a consequence of the symmetry of the entropy. With
$b=\cos n\phi$ the equilibrium equations are
\begin{equation}
    \label{eq:gleichrad}
    \frac{\partial}{\partial r} \tilde{s}(e, r, b) \stackrel{}{=} 0
\end{equation}
for the modulus of the  spontaneous magnetisation and
\begin{equation}
    \label{eq:gleichwink}
    \sin( n\phi) \frac{\partial }{\partial b} \tilde{s}(e, r, b)  \stackrel{}{=} 0
\end{equation}
for its angular dependence. The equilibrium condition for the
angular part can be satisfied in two ways. Either the derivative
$\partial_b \tilde{s}$ vanishes or $\sin n\phi$ is zero. The
latter is the case for $\phi = k\pi/n$ with $k = 0, \ldots, 2n-1$.
The former way to satisfy equation (\ref{eq:gleichwink}) is
discussed in section \ref{kap:energieabhwinkel}. The stability of
the solution of (\ref{eq:gleichrad}) and (\ref{eq:gleichwink})
requires that the corresponding Hessian is a negative definite
matrix. Suppose that the stability is guaranteed for the radial
order parameter component, i.e. $\partial_{rr}\tilde{s} < 0$. At
$\phi = k\pi/n$ the Hessian in polar coordinates reduces to
\begin{equation}
    H\left(e,\phi = k\pi/n\right) = \left(\begin{array}{cc}\frac{\partial^2}{\partial r^2} \tilde{s}& 0 \\
    0 & -\frac{n^2}{r^2}(-1)^k\frac{\partial}{\partial b} \tilde{s}    \end{array}\right)
\end{equation}
and hence the negativity of $H$ is satisfied if $
(-1)^k\partial_b\tilde{s} > 0$. The factor $(-1)^k$ can either be
$+1$ or $-1$ leading to maxima or saddle points of the
microcanonical entropy surface at the solutions of the equilibrium
conditions (\ref{eq:gleichrad}) and (\ref{eq:gleichwink}). For a
positive $\partial_b \tilde{s}$ the spontaneous magnetisation can
have the orientations $\phi = k\pi/n$ with even $k$. In this
situation the extrema on directions with odd $k$ are saddle points
of the entropy surface.

These considerations show that the direction of the spontaneous
magnetisation of an entropy with $C_{nv}$ invariance  is already
determined from the symmetry properties with respect to the
magnetisation. The extrema of the entropy appear along the
symmetry lines of the associated regular $n$-polygon in the
magnetisation plane.

To get further information about the behaviour of the entropy
surface for small magnetisations a Taylor expansion must actually
be performed. The precise form of this Taylor expansion, i.e. the
expansion coefficients and the degree of the truncated Taylor
polynomial determines the stability of the extrema and the form of
the energy dependence of the actual order parameter $r_{\scriptsize
\mbox{sp}}(e)=|m_{\scriptsize \mbox{sp}}(e)|$. To ensure the stability in
angular direction one must have $\partial_b\tilde{s}\neq 0$ and
hence the Taylor expansion has to contain at least the $n$th
degree monomial $r^n b=r^n \cos n\phi$.

\subsection{Entropy of the four-state vector Potts model}
\label{kap:vierpotts}

The general results of the previous subsection can be illustrated
for an entropy with $C_{4v}$ symmetry in the magnetisation space.
The rational basis of invariants is then given by $\{r^2, r^4 \cos
4\phi\}$ and the Taylor expansion of the entropy $s(e, r, \phi)$
up to the fourth degree term has the general form
\begin{equation}
    \label{eq:c4v_entwicklung}
    s(e,r,\phi) = s_0(e) + \frac{1}{2}A(e)r^2 + \frac{1}{4}B(e)r^4 +
    \frac{1}{4}C(e)r^4 \cos 4\phi \;.
\end{equation}
To ensure the appearance of a stable extremum for finite
magnetisations the coefficients of the fourth degree term have to
satisfy $B(e) + C(e)\cos 4\phi < 0$ for all angles $\phi$. The
extrema of the entropy are along the symmetry directions $\phi = k
\pi/4$ with $k = 0, \dots, 7$ of the square in the two-dimensional
plane. The stability in angular direction requires that the
inequality
\begin{equation}
    C(e) \cos k\pi > 0
\end{equation}
is satisfied. For a positive coefficient $C(e)$ this has the
consequence that the maxima of the microcanonical entropy surface
lie on the directions $\phi\in\{0, \pi/2, \pi, 3\pi/2\}$ provided
the stability of the radial order parameter is ensured. The
extrema on the four remaining symmetry directions correspond to
saddle points in the entropy function.

An example of a classical spin model with $C_{4v}$ symmetry is the
vector Potts model (Wu, 1982) with four states whose Hamiltonian is given by
\begin{equation}
    \label{eq:z4_hamilton}
  \mathcal{H} = -\sum_{\left<i,j\right>} \cos\left(\frac{\pi}{2}(k_i-k_j)\right) \;.
\end{equation}
The spins $k_i$ can take on the values 1, 2, 3 and 4 and be
visualised by unit vectors with the orientations $0$, $\pi/2$,
$\pi$ and $3\pi/2$ in a two-dimensional plane. The intensive
magnetisation of the system with $N$ spins is given by
\begin{eqnarray}
  m_1 & = & \frac{1}{N} \sum_i \cos\left( \frac{\pi}{2} k_i\right) \\
  m_2 & = & \frac{1}{N} \sum_i \sin\left( \frac{\pi}{2} k_i\right) \;.
\end{eqnarray}
The system has four equivalent ground states with the spontaneous
magnetisations $(1,0)$, $(0,1)$, $(-1,0)$ and $(0,-1)$. These four
ground state magnetisations define a square in the magnetisation
plane.

\begin{figure}
\begin{center}
\epsfig{file=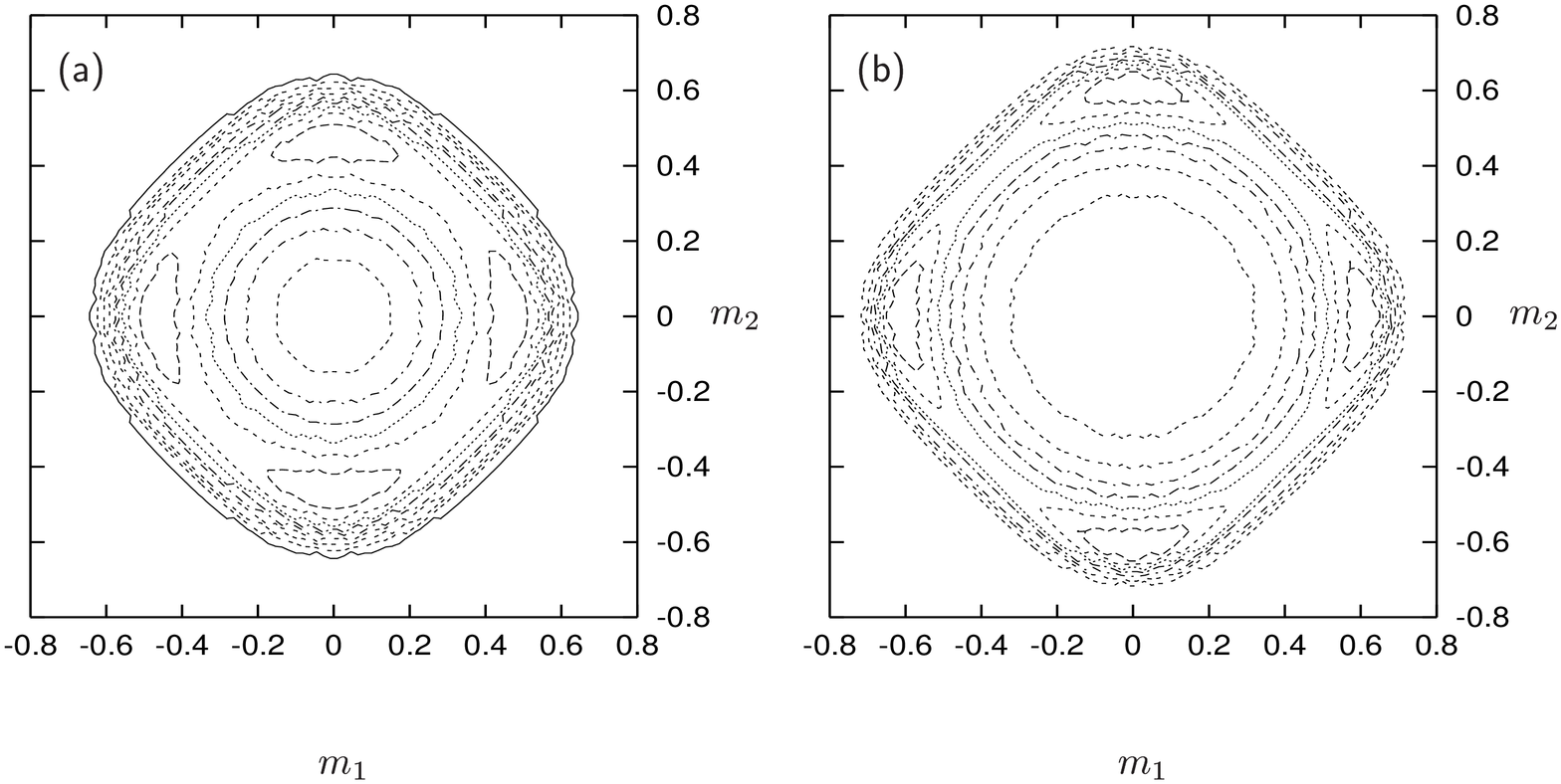,width=16em, angle=270}
\end{center}
\caption{\label{bildhoehen_z4} \small Level curves of the density
of states of the four-state vector Potts model with $64$ spins in
two dimensions for the two energies $-0.91$ (a) and $-1.06$ (b)
below the critical point $e_{\scriptsize \mbox{c}}=-0.7\pm 0.005$. Four equivalent
maxima appear along the bisectors of the square defined by the
ground state magnetisations.}
\end{figure}

\begin{figure}
\begin{center}
\epsfig{file=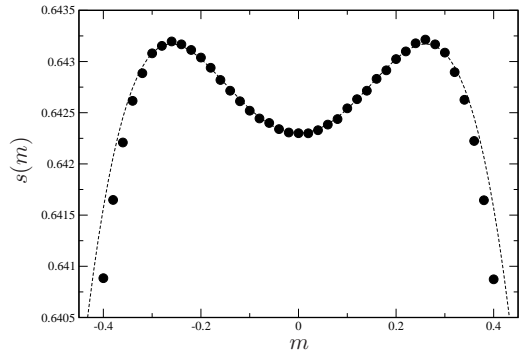,width=12em, angle=270}
\end{center}
\caption{\label{taylor_z4} \small The entropy of the two-dimensional four-state vector Potts model with $100$
spins at the energy $-0.86$. The filled circles display the simulated data along the
$m_1$ direction. The Taylor approximation
(\ref{eq:c4v_entwicklung}) up to the fourth degree term is shown as a dashed line. The second degree
coefficient $0.0257$ is obtained by differentiation, the fourth degree
coefficient $\frac{1}{4}(B+C) = -0.189$ by variation.
Additional higher order terms  are necessary to describe the behaviour of the entropy for
large magnetisations.}
\end{figure}

For a two-dimensional system with a finite number of  spins the
entropy has one single maximum at zero magnetisation above the
critical energy $e_{\scriptsize \mbox{c}}$. Below $e_{\scriptsize \mbox{c}}$ the
entropy has four equivalent maxima along the directions defined by
the coordinate axes. The extrema on the diagonals of the
coordinate system correspond to saddle points of the entropy
surface. In figure \ref{bildhoehen_z4} the level curves of the
entropy of a two-dimensional system with $64$ spins are shown for
two energies below the critical one. The data are calculated with
the transition observable method that allows a highly efficient
and accurate determination of microcanonical entropy surfaces
(H\"uller/Pleimling, 2002). The findings show that the angular
equilibrium equation  (\ref{eq:gleichwink}) is satisfied by the
condition $\sin 4\phi = 0$ for the four-state vector Potts model.
The appearance of stable phases along the coordinate axes suggests
that the coefficient $C(e)$ is indeed positive. However it is
hardly possible to obtain a reliable estimate of the fourth
derivative from an entropy that has been calculated by means of
Monte Carlo simulations. A direct determination of the expansion
coefficients $B(e)$ and $C(e)$ is therefore not possible. To get
an impression of the values of these coefficients one can use an
indirect approach. The coefficient $A(e)$ of the second degree
term in the expansion (\ref{eq:c4v_entwicklung}) is determined by
differentiating the entropy. The coefficients $B(e) \pm C(e)$
corresponding to the direction $0$ and $\pi/2$ are then varied to
achieve a good description of the data by the fourth degree
expansion for small magnetisations. For the energy $-0.75$ near
$e_{\scriptsize \mbox{c}}=-0.7$ this gives for the system with $L=8$ the
value $-1.14$ for $B$ and the positive value $0.04$ for $C$ in
agreement with the general considerations above. For a system with
$L=10$ this procedure yields the coefficients $B=-0.804$ and the
positive value $0.048$ for $C$ at the energy $-0.86$ below the
critical energy $-0.806\pm 0.005$. Note however that this
procedure may lead to large errors due to the subjective
assessment of the agreement of the Taylor polynomial with the data
for small magnetisations. The simulated entropy for the system
with $100$ spins together with its fourth degree Taylor
approximation along the $m_1$ direction is displayed in figure
\ref{taylor_z4}.

\begin{figure}
\begin{center}
\epsfig{file=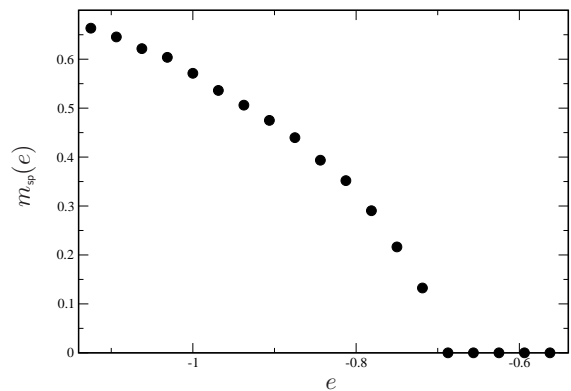,width=13em, angle=270}
\end{center}
\caption{\label{bildsponmag_z4} \small The spontaneous magnetisation of the four-state vector Potts model
with $64$ spins in two dimensions as a function of the energy.
The data suggest a critical energy $e_{\scriptsize \mbox{c}} = -0.7\pm 0.005$.}
\end{figure}

\begin{figure}
\begin{center}
\epsfig{file=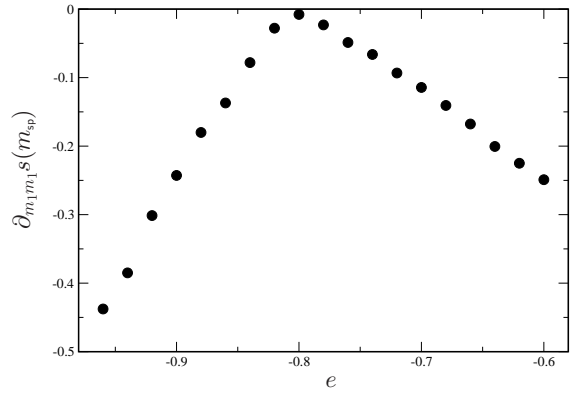,width=13em, angle=270}
\end{center}
\caption{\label{bildsmm_z4} \small The (equilibrium) curvature of
the entropy surface of the two-dimensional vector Potts model with
$100$ spins along the $m_1$ direction. Near the critical point
$e_{\scriptsize \mbox{c}}=-0.806\pm 0.005$ the curvature vanishes leading to a
divergent susceptibility. At the critical point the moduli of the
gradients from the left and from the right differ by the factor
$1.98$ in good agreement with the prediction from the Taylor
expansion.}
\end{figure}

\begin{figure}
\begin{center}
\epsfig{file=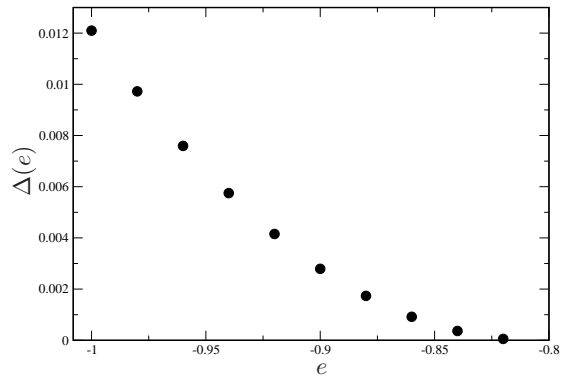,width=13em, angle=270}
\caption{\label{bilddurch_z4} \small The depth of the convex dip
of entropy surface of the two-dimensional vector Potts model with
$100$ spins along the $m_1$ direction. Near the critical point
$e_{\scriptsize \mbox{c}}=-0.806\pm 0.005$ the energy variation is most suitably
described by a parabola.}
\end{center}
\end{figure}

The behaviour of the spontaneous magnetisation of the
two-dimensional system with $64$ spins resulting from the
variation of the position of the maximum is displayed in figure
\ref{bildsponmag_z4}. Its energy dependence  near the critical
energy $e_{\scriptsize \mbox{c}}=-0.7$ is most suitably characterised by a
square root behaviour, i.e. by the exponent $\tilde{\beta} = 1/2$.
This can be understood from the general considerations in section
\ref{kap:taylor}. The non-zero spontaneous magnetisation
$m_{sp}(e)$ defines a vector in the order parameter space with a
energy-dependent modulus $|m_{sp}(e)|$ and a fixed,
energy-independent direction $m_0$ (compare relation
(\ref{eq:zerlegung_ord})). Once the direction $m_0$ has been
chosen, e.g. along the $m_1$ axis, the entropy depends only on the
modulus $|m|$ leading to an entropy function with a
one-dimensional order parameter. This function is just the cut of
the entropy surface along the $m_0$ direction. The expansion of
this function with respect to $|m|$ can be performed as in section
\ref{kap:taylor} and yields the characteristic asymptotic square
root energy behaviour of the microcanonical equilibrium
magnetisation  of finite systems near the transition energy
$e_{\scriptsize \mbox{c}}$. The curvature parallel to the spontaneous
magnetisation at the equilibrium macrostate of the system with
$100$ spins near the critical point is shown in figure
\ref{bildsmm_z4}. It results in a diverging susceptibility
characterised by the critical exponent $\tilde{\gamma} = 1$. The
amplitudes from the left and from the right differ approximately
by a factor $2$. These observations are in agreement with the
general findings obtained by Taylor expanding the microcanonical
entropy surface. Note however that the susceptibility of the
four-state vector Potts model is a tensor. Due to the symmetry
(\ref{eq:entro_symmetrie_c4n}) of the entropy surface the
susceptibility tensor is diagonal with a non-vanishing parallel
and perpendicular component with respect to the equilibrium
magnetisation. The behaviour of the depth of the convex dip of the
entropy surface below the critical energy $e_{\scriptsize \mbox{c}}$ is
shown in figure \ref{bilddurch_z4} for the system with $100$
spins. Once again the parabolic energy dependence suggested by the
above considerations is confirmed.

\subsection{Energy-dependent angular magnetisation}
\label{kap:energieabhwinkel}

To complete the discussion of the angular equilibrium equation
(\ref{eq:gleichwink}) this section focuses on the possible
solution $\partial_b \tilde{s} = 0$. The derivative $ \partial_b
\tilde{s}$ contains the radial magnetisation so that the two
equilibrium equations for $r_{\scriptsize \mbox{sp}}$ and $\phi_{\scriptsize
\mbox{sp}}$ are coupled.  This set of equations can be solved for
$b$ and gives $b_{\scriptsize \mbox{sp}}$ or equivalently $\phi_{\scriptsize
\mbox{sp}}$ as a function of $r_{\scriptsize \mbox{sp}}$ and hence it will
depend on the energy. Consequently The direction of the
spontaneous magnetisation exhibits an energy dependence. This
dependence on the energy is influenced by the precise form of the
truncated Taylor expansion which determines the order parameter as
a function of the energy. The stability of the extremum requires
additionally that the expansion must contain monomials up to the
degree $2n$ at least. To see this consider the component
$H_{\phi\phi}$ of the Hessian which is given by
\begin{equation}
    H_{\phi\phi}^{(0)} = \frac{n^2}{r_{\scriptsize \mbox{sp}}^2} (\sin n\phi_{\scriptsize
    \mbox{sp}})^2 \partial_{bb} \tilde{s}
\end{equation}
at the spontaneous magnetisation. This requirements for the Taylor
expansion has the further consequence that the phase diagram with
respect to the expansion coefficients of the entropy allows only
an isolated continuous transition (Gufan/Sakhnenko, 1973).

\begin{figure}
\begin{center}
\epsfig{file=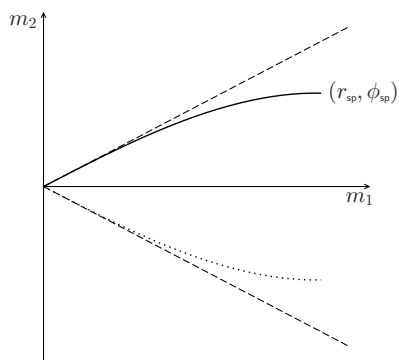,width=12em, angle=270}
\end{center}
\caption{\label{winkelenerabh} \small Schematic behaviour of the magnetisation curve
(solid line) $(r_{\scriptsize \mbox{sp}}, \phi_{\scriptsize \mbox{sp}})$ in the
two-dimensional plane that results from the model entropy
(\ref{eq:diemodellentropie}). The angles $-\pi/8$ and $\pi/8$ are
indicated by dashed lines. A second solution (dotted line) for the
order parameter is obtained by reflecting the spontaneous
magnetisation about the $m_1$ axis. Application of the rotations
in $C_{4v}$ to these two solutions gives the remaining solutions
for the order parameter.}
\end{figure}

The author does not know of any model system that exhibits an energy-dependent
orientation of the microcanonically defined order parameter. Nevertheless,
the general statements of this section can be illustrated with a
model entropy which is invariant under the symmetry group
$C_{4v}$. With $b = \cos 4\phi$ and $\varepsilon = e - e_{\scriptsize
\mbox{c}}$ the model entropy to investigate is given by
\begin{equation}
\label{eq:diemodellentropie}
    \tilde{s}(\varepsilon,r,b) = s_0(\varepsilon) -
    \frac{A}{2}\varepsilon r^2
    -\frac{B}{4}r^4 + \frac{C}{4}\varepsilon^3br^4
    -\frac{G}{8}b^2r^8
\end{equation}
with positive coefficients $A$, $B$ and $G$ below $e_{\scriptsize
\mbox{c}}$. The coefficient $C$ has to be non-zero. For the
following discussion it is  assumed to be negative. The
equilibrium equations for a non-zero spontaneous magnetisation
$(r_{\scriptsize \mbox{sp}}, \phi_{\scriptsize \mbox{sp}})$ are
\begin{equation}
\label{eq:erstegleichung_e}
    -A\varepsilon -Br^2 + C\varepsilon^3 br^2  - Gb^2r^6=0
\end{equation}
and
\begin{equation}
\label{eq:zweitegleichung_e}
    \sin 4\phi \left( C\varepsilon^3r^4 - Gbr^8 \right)=0 \;.
\end{equation}
The second equation (\ref{eq:zweitegleichung_e}) can be solved for
$b$ which yields
\begin{equation}
\label{eq:bausdruck}
    b = \frac{C\varepsilon^3}{Gr^4}\;.
\end{equation}
At this stage it is already obvious that the direction of the
magnetisation vector in equilibrium will exhibit a dependence on
the energy. Plugging this result into (\ref{eq:erstegleichung_e})
gives rise to the expression
\begin{equation}
    r_{\scriptsize \mbox{sp}} = \sqrt{\frac{A}{B}|\varepsilon|}
\end{equation}
for the radial order parameter. In view of equation
(\ref{eq:bausdruck}) one obtains
\begin{equation}
\label{eq:zuinvertieren}
    b_{\scriptsize \mbox{sp}} = \cos 4 \phi_{\scriptsize \mbox{sp}} =
    \frac{CB^2}{GA^2}\varepsilon \;.
\end{equation}
Inverting this result for an angle $\phi_{\scriptsize \mbox{sp}} \in
[0,\pi/4]$ one ends up with the energy-dependent spontaneous
direction
\begin{equation}
    \phi_{\scriptsize \mbox{sp}}(\varepsilon) = \frac{1}{4} \arccos \left(
    \frac{CB^2}{GA^2} \varepsilon\right)\;.
\end{equation}
Note that a different interval for the angle has to be chosen for
a positive coefficient $C$. The resulting curve $(r_{\scriptsize
\mbox{sp}}(\varepsilon), \phi_{\scriptsize \mbox{sp}}(\varepsilon))$ in the
two-dimensional $m_1m_2$-plane is schematically depicted in figure
\ref{winkelenerabh}.
Only one solution for the angle of the spontaneous magnetisation
has been considered so far. As the entropy surface
(\ref{eq:diemodellentropie}) is invariant under the group $C_{4v}$
the other solutions can be obtained by applying the
transformations of $C_{4v}$ onto the solution $(r_{\scriptsize \mbox{sp}},
\phi_{\scriptsize \mbox{sp}})$ from above. This results in a star of eight
possible order parameters with an energy-dependent direction. The
appearance of eight solutions is related to the eight different
choices of the interval of length $\pi$ for the values of $4\phi$
in the interval $[0,8\pi]$ when inverting relation
(\ref{eq:zuinvertieren}). The Hessian of the model entropy
evaluated at the solution for the order parameter is diagonal. The
stability of the solution requires that the this Hessian has
negative diagonal entries. This is indeed the case as
\begin{equation}
H_{rr}^{(0)} = -2A|\varepsilon| - \frac{4C^2}{G} \epsilon^6
\end{equation}
and
\begin{equation}
    H^{(0)}_{\phi\phi} = -4 (\sin 4 \phi_{\scriptsize \mbox{sp}})^2 Gr_{\scriptsize \mbox{sp}}^6
\end{equation}
are both negative.

\section{Summary}

The appearance of classical critical exponents in physical
observables of finite microcanonical systems such as the order
parameter and the susceptibility is related to the assumed
analyticity of the microcanonical entropy of finite systems. This
can be demonstrated by Taylor expanding the entropy function about
the well-defined critical point of the finite microcanonical
system. The general findings are confirmed by concrete
investigations of spin systems. This suggests in turn that the
assumption of an analytic microcanonical entropy function is
indeed justified although a proof is still lacking. An interesting
example of a classical spin system is the four-state vector Potts
model with a two-component order parameter. The form of the Taylor
expansion of the entropy of the four-state vector Potts model is
determined by the order parameter symmetry. The general
consequences of the $C_{4v}$ symmetry on the form of the expansion
of the microcanonical entropy and the resulting behaviour of the
physical quantities for small magnetisations can be nicely
verified by Monte Carlo calculations. The analyticity together
with the knowledge of the order parameter symmetry allows already
far-reaching statements about the physics of finite systems in the
microcanonical ensemble.

\ack

The author is much obliged to Efim
Kats for his hospitality at the Institut Laue-Langevin in Grenoble where
most of this work was carried out.
He would like to thank Alfred H\"uller and Michel Pleimling as well for
stimulating discussions and helpful comments on the manuscript.
The data of figure 1 was kindly supplied by Alfred H\"uller.

\section*{References}

\begin{harvard}

\item[]{Barr\'{e} J, Mukamel D, and  Ruffo S} 2001, {
Inequivalence of Ensembles in a System with Long-Range Interactions},
{\em Phys. Rev. Lett.} {\bf 87}, 030601

\item[]{Behringer H} 2003, {Symmetries of Microcanonical Entropy
Surfaces}, {\em J. Phys. A: Math. Gen.} {\bf 36}, 8739

\item[]{Binder K} 1972, {Statistical Mechanics of Finite Three-Dimensional Ising Models},
Physica {\bf 62}, 508

\item[]{Creswick R\,J } 1995, {Transfer Matrix for the Restricted
Canonical and Microcanonical Ensembles}, {\em Phys. Rev. E} {\bf 52},
5735

\item[]{Dauxois T, Holdsworth P, and Ruffo S} 2000, {Violation of Ensemble Equivalence in the
Antiferromagnetic Mean-Field XY Model}, {\em Eur. Phys. J. B} {\bf
16}, 659

\item[]{Gross D\,H\,E} 1986a, On the Decay of Very Hot Nuclei: Microcanonical Metropolis
Sampling of Multifragmentation, in W\,U Schr\"oder, ed, 1986, {\em
Nuclear Fission and Heavy Ion Induced Reactions}, p 427
(Rochester)

\item[]{Gross D\,H\,E} 1986b, On the Decay of Highly Excited Nuclei into Many
Fragments --- Microcanonical Monte Carlo Sampling, in M di Toro et
al., eds, 1986, {\em Topical Meeting on Phase Space Approach to
Nuclear Dynamics, Triest 1985}, p 251 (Singapore: World
Scientific)

\item[]{Gross  D\,H\,E} 2001, {\em Microcanonical Thermodynamics:
Phase Transitions in ``Small'' Systems (Lecture Notes in Phyiscs 66)}, (Singapore:World Scientific)

\item[]{Gross D\,H\,E, Ecker A, and Zhang X\,Z} 1996, {
Microcanonical Thermodynamics of First Order Transitions Studied
in the Potts Model}, {\em Ann. Phys.} {\bf 5}, 446

\item[]{Gufan Yu\,M} 1971,  {Phase Transitions Characterized by a Multicomponent
Order Parameter}, {\em Sov. Phys, Solid State} {\bf 13}, 175

\item[]{Gufan Yu\,M,  and Sakhnenko V\,P} 1973, { Features of
Phase Transitions Associated with Two- and Three-Component Order
Parameters}, {\em Sov. Phys. JETP} {\bf 36}, 1009

\item[]{H\"uller A} 1994, {First Order Phase Transitions in the
Canonical and the Microcanonical Ensemble }, {\em Z.  Phys. B} {\bf 93},
401

\item[]{H\"uller A,  and Pleimling M} 2002, {Microcanonical
Determination of the Order Parameter Critical Exponent}, {\em Int. J.
Mod. Phys. C} {\bf 13}, 947

\item[]{Ispolatov I,  and  Cohen E\,G\,D.} 2001, { On First-Order
Phase Transitions in Microcanonical and Canonical Non-Extensive
Systems}, {\em Physica A} {\bf 295}, 475

\item[]{Kastner M} 2002, {Existence and Order of the Phase
Transition of the Ising Model with Fixed Magnetization }, {\em J. Stat.
Phys.} {\bf 107}, 133

\item[]{Kastner M,  Promberger M, and H\"uller A} 2000,
{Microcanonical Finite-Size Scaling}, {\em J. Stat. Phys.} {\bf 99},
1251

\item[]{Lee T\,D and Yang C\,N} 1952, {Statistical Theory of
Equations of State and Phase Transitions. II. Lattice Gas and
Ising Model}, {\em Phys. Rev.} {\bf 82}, 410

\item[]{Lewis J\,T,   Pfister C-E, and Sullivan W\,G} 1994,
{The Equivalence of Ensembles for Lattice Systems: Some Examples
and a Counterexample}, {\em J. Stat. Phys.} {\bf 77}, 397

\item[]{Promberger M,  and H\"uller A} 1995, {Microcanonical
Analysis of a Finite Three-Dimensional Ising System}, {\em Z. Phys. B}
{\bf 97}, 341

\item[]{Richer A, Pleimling M, H\"uller A 2004, to be published}

\item[]{Stanley H\,E} 1972, {\em Introduction to Phase Transitions
and Critical Phenomena}, (Oxford: Oxford University Press)

\item[]{Tol\'edano J-P, and Tol\'edano P} 1987, {\em The Landau Theory of Phase Transitions}, (Singapore: World Scientific)

\item[]{Wu F\,Y} 1982, {The Potts Model}, {\em Rev. Mod. Phys.} {\bf
54}, 235

\item[]{Yang C\,N,  and Lee T\,D} 1952, {Statistical Theory of
Equations of State and Phase Transitions. I. Theory of
Condensation}, {\em Phys. Rev} {\bf 87}, 404

\end{harvard}

\Figures

\newpage

\end{document}